\documentclass[letterpaper,twocolumn,10pt]{article}
\usepackage{usenix-2020-09}

\usepackage[utf8]{inputenc}
\usepackage{url}
\usepackage{amsmath}
\usepackage{amssymb}
\usepackage{graphicx}
\usepackage{array}
\usepackage[english]{babel}
\usepackage{threeparttablex}
\usepackage{multirow}
\usepackage{pifont}
\usepackage{multicol}
\usepackage{caption}
\usepackage{rotating}
\usepackage[lofdepth,lotdepth]{subfig}

\newcommand*\rot{\rotatebox{90}}

\begin{document}

\title{\Large \bf {Detecting Malicious Accounts showing Adversarial Behavior in Permissionless Blockchains}

 \author{
 {\rm Rachit Agarwal}\\
 IIT-Kanpur
 \and
 {\rm Tanmay Thapliyal}\\
 IIT-Kanpur
 \and
 {Sandeep K Shukla}\\
 IIT-Kanpur
 }
 }

\maketitle
\begin{abstract} 

Different types of malicious activities have been flagged in multiple permissionless blockchains such as bitcoin, Ethereum etc. While some malicious activities exploit vulnerabilities in the infrastructure of the blockchain, some target its users through social engineering techniques. To address these problems, we aim at automatically flagging blockchain accounts that originate such malicious exploitation of accounts of other participants. To that end, we identify a robust supervised machine learning (ML) algorithm that is resistant to any bias induced by an over representation of certain malicious activity in the available dataset, as well as is  robust against adversarial attacks. We find that most of the malicious activities reported thus far, for example, in Ethereum blockchain ecosystem, behaves statistically similar. Further, the previously used  ML algorithms for identifying malicious accounts show bias towards a particular malicious activity which is over-represented. In the sequel, we identify that Neural Networks (NN) holds up the best in the face of such bias inducing dataset at the same time being  robust against certain adversarial attacks. 
\end{abstract} 

\section{Introduction}\label{sec:intro}

Blockchains can be modeled as ever-growing temporal graphs, where interactions (also called as \textit{transactions}) happen between different entities. In a blockchain, various transactions are grouped to form a \textit{block}. These blocks are then connected together to form a blockchain. A typical blockchain is immutable and is characterized by properties such as confidentiality, anonymity, and non-repudiability. Irrespective of the type of blockchain (which are explained next), these properties achieve a certain level of privacy and security. There are mainly two types of blockchains - permissionless and permissioned blockchains. In permissioned blockchains, all actions on the blockchain are authenticated and authorized, while in permissionless blockchains such aspects are not required for successful transactions. Permissionless blockchains usually also support a native crypto-currency.

Lots of malicious activities, such as ransomware payments in crypto-currency and Ponzi schemes~\cite{Bartoletti2018}, happen due to the misuse of the permissionless blockchain platform. A malicious activity is where accounts perform illegal activities such as Phishing, Scamming and Gambling. In~\cite{Maleh2020}, the authors survey different types of attacks and group them based on the vulnerabilities they target in the permissionless blockchain. Thus, one question we ask is \textit{can we train a machine learning algorithm (ML) to detect malicious activity and generate alerts for other accounts to safeguard them?} There are various state-of-the-art approaches, such as~\cite{agarwal2020detecting, pham2016anomaly, Zola2019}, that use ML algorithms to detect malicious accounts in various permissionless blockchains such as Ethereum~\cite{Vitalik2018} and Bitcoin~\cite{nakamotobitcoin}. In~\cite{agarwal2020detecting}, the authors outline an approach to detect malicious accounts where they consider the temporal aspects inherent in a permissionless blockchain including the transaction based features that were used in other related works such as~\cite{pham2016anomaly, Zola2019}. 

Nonetheless, these related works have various drawbacks. They only study whether an account in the blockchain is malicious or not. They do not classify or comment upon the type of malicious activity (such as Phishing or Scamming) the accounts are involved in. Further, as they do not consider the type of malicious activities, they do not study the bias induced by the imbalance in the number of accounts associated with  particular malicious activities and, thus, fail to detect differences in  the performance of the identified algorithm on the different kinds  of malicious activities. Additionally, they do not study the performance of the identified algorithm when any adversarial input is provided. An adversarial input is defined as an intelligently crafted account whose features hide the malicious characteristics that an ML algorithm uses to detect malicious accounts. Such accounts may be designed to evade ML based detection of their maliciousness. 

Thus, we are motivated to answer: (\textbf{Q1}) \textit{can we effectively detect malicious accounts that are  represented in minority in the blockchain?} Also, we consider another question and answer: (\textbf{Q2}) \textit{can we effectively detect malicious accounts that show  behavior adversarial yp ML based detection?} We answer these two questions using a three-fold methodology. \textbf{\textit{(R1)}} First, we analyze the \textit{\textbf{similarity}} between different types of malicious activities that are currently known to exist. Here we understand if under-represented malicious activities do have any similarity with other malicious activities. \textbf{\textit{(R2)}} We then study the \textit{\textbf{effect of bias}} induced in the ML algorithm, if any, by the malicious accounts attributed to a particular malicious activity which is represented in large number. We then identify the ML algorithm which we can use to efficiently detect not only the largely represented malicious activities on the blockchain, but also activities that are under-represented. For the state-of-the-art approaches, we train and test different ML algorithms on the dataset and compute the recall considering all malicious activities under one class. Next, to understand the robustness of the identified ML algorithm, we test it on the malicious accounts that are newly tagged with a motivation to understand, if in future, such accounts can be captured. \textbf{\textit{(R3)}} Nonetheless, we also test the robustness of the ML algorithm on the adversarial inputs. Here, we use Generative Adversarial Networks (\textbf{\textit{GAN}})~\cite{goodfellow2014} to first generate adversarial data using already known feature vectors of malicious accounts and then perform tests on such data using the identified ML model to \textit{\textbf{study the effect of adversarial attacks}}.

To facilitate our work, we focus on a permissionless blockchain called Ethereum blockchain~\cite{Vitalik2018}. Ethereum has mainly two types of accounts - Externally Owned Accounts (\textit{EOA}) and Smart Contracts (\textit{SC}) - where transactions involving EOAs are recorded in the ledger while transactions between SCs are not. There are various vulnerabilities in Ethereum~\cite{chen2019survey} that attackers exploit to siphon off the \textit{Ethers} (a crypto-currency used by Ethereum). Our study is not exhaustive for all possible attacks in a blockchain. However, we study only those for which we have example accounts. Note that our work is applicable to other permissionless blockchains as well that focus on transactions of crypto-currency. We choose Ethereum because of the volume, velocity, variety, and veracity within the transaction data. Currently, Ethereum has more than 14 different types of known fraudulent activities.

Our results show that certain malicious activities behave similarly when expressed in terms of feature vectors that leverage temporal behavior of the blockchain. We also observe that Neural Network (\textbf{NN}) performs relatively better than other supervised ML algorithms and the volume of accounts attributed to a particular malicious activity does not induce a bias in the results of NN, contrary to other supervised ML algorithms. Moreover, when an adversarial data is encountered, NN's performance is still better than the other supervised ML algorithms. Henceforth, note that when we refer to the performance of an ML algorithm we mean recall achieved on a testing dataset. 

In summary, our contributions, in this paper are as follows:

\begin{enumerate}
\item The similarity scores reveal that there \textbf{exists similarity} between different types of malicious activities present until 7th Dec 2019 in the Ethereum blockchain and they can be clustered in mostly 3 clusters. 

\item All the state-of-the-art ML algorithms used in the related studies \textbf{are biased towards} the dominant malicious activity, i.e., `Phishing' in our case. 

\item We identify that a \textbf{Neural Network} based ML model is the least affected by the imbalance in the numbers of different types of the malicious activities. Further, when adversarial input of the transactional data in the Ethereum blockchain is provided as test data, NN model is resistant to it. When NN is trained on some adversarial inputs, its balanced accuracy increases by 1.5\%. When trained with adversarial data, most other algorithms also regain their performance with RandomForest leading having the best recall. 

\end{enumerate}

The rest of the paper is organized as follows. In section~\ref{sec:back}, we present the background and the state-of-the-art techniques used to detect malicious accounts in blockchains. In sections~\ref{sec:method}, we present a detailed description of our methodology. This is followed by an in-depth evaluation along with the results in section~\ref{sec:eval}. We finally conclude in section~\ref{sec:conclusion} providing details on prospective future work.

\section{Background and Related Work}\label{sec:back}

Between 2011 and 2019 there have been more than 65 exploit incidents on various blockchains~\cite{Maleh2020}. These attacks mainly exploit the vulnerabilities that are present in the consensus mechanism, client, network, and  mining pools. For example, Sybil attack, Denial of Service attack, and the 51\% attack. In specific cases such as Ethereum, SC based vulnerabilities are also exploited. For instance, the Decentralized Autonomous Organization (DAO) attack\footnote{Understanding dao hack:  \url{https://www.coindesk.com/understanding-dao-hack-journalists}}, exploits the Reentrancy bug~\cite{Samreen2020} present in the SC of the DAO system. While some of the attacks exploit bugs and vulnerabilities, some exploits target users of the blockchain. The users are sometimes not well-versed with the technical aspects of the blockchain while sometimes they get easily influenced by the various social engineering attempts. Such exploits are also present in permissionless blockchains such as Ethereum. From the social engineering perspective, \textit{Phishing} is the most common malicious activity present in Ethereum, where it is represented by more than 3000 accounts~\cite{etherscanLC}. The Table~\ref{tab:MaliciousActivities} presents different known malicious activities that are reported to exist in the Ethereum blockchain and are used by us in this work. Note that some activities have similar description but are marked differently in Ethereum.

\begin{table*}[t]
    \centering
        \begin{tabular}{|r|l|}
        \hline
            \textbf{Malicious Incident/Activity}& {\textbf{Description}}\\
            \hline
           Lendf.Me Protocol Hack           & Exploit of a reentrancy vulnerability arising due to usage of ERC-777 token~\cite{lendf}.\\
           \hline
           EtherDelta Website Spoof         & Attackers spoofed the official EtherDelta website so that users transact through it~\cite{EtherDelta}.\\
           \hline
           \multirow{2}{*}{Ponzi Scheme}    & An attacker enticed a user to lend him crypto-currency, which the attacker \\
                & used to repay the debt of previously scammed user.\\
            \hline
           Parity Bug                       & Bug in a multi-sig parity wallet which caused freezing of assets.\\
           \hline
           \multirow{2}{*}{Phishing}                        & When attackers pose as legitimate to lure individuals into transacting \\
                                            & with them, example Bancor Hack~\cite{bancor}.\\
           \hline
           Gambling                         & Accounts involved in Gambling activities which is illegal in many countries.\\
           \hline
           Plus Token Scam                  & A Ponzi scheme~\cite{plusToken}.\\
           \hline
           Compromised & Accounts whose address were either compromised or were scammed.\\
           \hline
           Scamming     & Accounts which are reported to be involved in frauds.\\
           \hline
           \multirow{2}{*}{Cryptopia Hack}  & No official description, but unauthorized and unnoticed transfer happened from \\
                                            & Cryptopia wallet to other exchanges~\cite{Cryptopia}.\\
            \hline
           Bitpoint hack                    & ``\textit{Irregular withdrawal from Bitpoint exchange's hot wallet}''~\cite{BITPOINT}.\\
           \hline
           Fake Initial Coin Offerings      & Fake Startups aimed to siphon off crowd-funded investments.\\
           \hline
           \multirow{2}{*}{Upbit Hack} & Speculation that an insider carried out malicious activity when the exchange was  \\
           & moving funds from hot to cold wallet~\cite{upbit}.\\
           \hline
           Heist & Account involved in various Hacks such as Bitpoint Hack.\\
           \hline
           Spam Token & No official description on the activity.\\
           \hline
           Suspicious & No official description but account are mainly involved in various scams.\\
           \hline
           Scam & No official description but account are mainly involved in various scams.\\
           \hline
           Unsafe & No official description but account are mainly involved in various scams.\\
           \hline
           Bugs & Accounts whose associated activities caused issues in the system\\
           \hline
        \end{tabular}
    \caption{Malicious Activities on Ethereum blockchain as reported by Etherscan.}
    \label{tab:MaliciousActivities}
\end{table*}

Variety of techniques and approaches have been used to detect, study and mitigate such different types of attacks and vulnerabilities in blockchains. In~\cite{chen2019survey}, authors survey different vulnerabilities and attacks in Ethereum blockchain and provide a discussion on different defenses employed. We classify these defenses into 3 groups: (a) those that deploy honeypots to capture and analyse transactions in the blockchain, (b) those that use wide variety of machine learning algorithms to analyse transactions, and (c) those that study the vulnerabilities in the system such as bytecode of the smart contracts to analyse malicious activities. 

In~\cite{cheng2019}, the authors deployed a honeypot and studied the attacks that happen in the Ethereum blockchain. They analyzed the HTTP and Remote Procedure Call (RPC) requests made to the honeypot and performed behavioral analysis of the transactions. They found adversaries follow specific patterns to steal crypto-currency from the blockchain. Nonetheless, in some instances such honeypots are also compromised, for example the honeypot at the address - `\texttt{0x2f30ff3428d62748a1d993f2cc6c9b55df40b4d7}'.

In~\cite{agarwal2020detecting}, the authors present a survey of different state-of-the-art ML algorithms that are used to detect malicious accounts in a blockchain transaction network and then presented the need for the temporal aspects present in blockchains as new features. Here, we refrain ourselves from surveying again the methods already presented in~\cite{agarwal2020detecting}. Instead, we present their findings and new techniques that have been reported since. In~\cite{agarwal2020detecting}, the authors categorized the features into two main categories: transaction based and graph based. With respect to the transaction based features, they reported the use of features such as \textit{Balance In}, \textit{Balance Out}, and \textit{Active Duration}. With respect to the graph based features, they identified the extensive use of the features such as clustering coefficient~\cite{Watts1998} and in/out-degree. The authors, motivated to capture the diversity in the transactions, found that the use of temporal features further enhanced the robustness of the ML algorithm used towards identifying malicious accounts in a blockchain. These features amounted to a total of 59 features and were related to inter-event transaction properties such as the stability of neighborhood (referred to as \textit{attractiveness}) and non-uniformity (referred to as \textit{burst}~\cite{Karsai2012}) present in the degree, inter-event time, gas-price, and balance. A complete list of feature used in~\cite{agarwal2020detecting} is presented in appendix~\ref{app:b}. Using such enriched feature vector, they validated their approach and achieved high recall ($>78\%$) on the entire class of malicious accounts present in their test dataset. 

In~\cite{Ala2020}, the authors used Graph Convolutional Networks (GCN) to detect money-laundering related malicious activities in the Bitcoin blockchain. They developed a Bitcoin transaction graph where the transactions were represented as the nodes while the flow of Bitcoin (crypto-currency used in Bitcoin blockchain) was represented as  the edges. They used transaction based features such as amount of Bitcoins received and spent by an account and the Bitcoin fee incurred by a transaction. Using GCN, they achieved F1 score of 0.77 on their dataset. Similarly, in~\cite{lorenz2020machine}, the authors constructed a transaction graph with similar features as in~\cite{Ala2020} to detect malicious activities in the Bitcoin blockchain. They compared the use of unsupervised, supervised and active learning approaches. They observe that the Active Learning techniques performed better.  

In~\cite{wuJ2019} the authors explored the transactions carried out by different accounts that were involved in Phishing activity. They analyzed the transaction data and proposed \textit{trans2vec}, a network embedding algorithm, to extract features from the Ethereum blockchain data. They, then, used the extracted features with One Class Support Vector Machine (OC-SVM) to detect accounts involved in phishing activities, and achieved a recall score of 89.3\% on the malicious class. Although, they focused on the phishing activities, they did not discuss the applicability of \textit{trans2vec} with respect to other types of malicious activities.

In~\cite{Zhang2020}, the authors developed a framework to analyze the transactions on the Ethereum blockchain and detect various attacks which exploit different vulnerabilities therein. They replayed all transactions related to a particular address and monitor the Ethereum Virtual Machine (EVM) state. They, then, applied logic rules on the transactions to detect abnormal behavior associated with a particular vulnerability and study only \textit{Suicidal}, \textit{UncheckedCall}, and \textit{Reentrancy} vulnerabilities. 

In all the above-mentioned work, the authors did not distinguish between the behaviors of different types of malicious activities that are present in permissionless blockchains. Further, no discussion is provided on the effect of adversarial data on the approaches. To the best of our knowledge, in the field of blockchain security, the current work is the first instance that studies data poisoning and evasion and tries to safeguard them against any adversarial attack.

As adversarial data might not be present in the dataset, \textbf{\textit{GAN}}~\cite{goodfellow2014} is one of the most commonly used techniques to generate adversarial data for testing the approach. Based on Neural Network, GANs were originally proposed for usage in the field of computer vision and machine translation, but over the years the technique has gained popularity in various sub-domains of cyber-security such as intrusion detection systems~\cite{chen2019efficient}. The architecture of GAN consists of two basic components: a generator and a discriminator. A generator generates fake data that has similar characteristics as the original data. The fake generated data along with the real data is passed to the discriminator which discriminates the input data and identifies if it is real or fake. Both the generator and the discriminator are trained iteratively over the dataset. Over time, generator becomes more intelligent, making it hard for the discriminator to correctly classify the real and the fake data. There are many variants of GANs that are available\footnote{Different GAN algorithms:  \url{https://github.com/hindupuravinash/the-gan-zoo}}. We do not describe all the techniques and variants of GAN as this is out of the scope of this work. However, here, we only describe CTGan~\cite{xu2019} that we use to generate fake data for every malicious activity represented by accounts in our dataset. Our choice of CTGan is based on the fact that CTGan is able to generate tabular data. In CTGan, the generator and discriminator models contain 2 fully connected hidden layers, which account for capturing correlation in the feature vectors given to them. They use Rectified Linear Units (ReLU) activation function along with batch normalization in generator model to reduce over-fitting. ReLU is defined as the positive part of the argument passed to activate the neuron. The discriminator has leaky ReLU activation along with dropout regularization~\cite{srivastava2014} implemented in each hidden layer. When using CTGan for data generation, for the best results, the authors recommend the number of epochs (where one epoch represents single iteration over the dataset) to be greater than 300. One limitation of CTGan is that the generator model needs at least 10 feature vectors to generate adversarial data. 

\section{Our Approach}\label{sec:method}

In this section, we describe our three-fold approach towards answering our research questions, in detail. 

\subsection{Computing similarity amongst malicious accounts}
 
We compute cosine similarity measure amongst accounts attributed to different known malicious activities to understand if the malicious activities have similar behavior. We acknowledge that there are other methods to quantify similarity, but in this work we use cosine similarity as it is widely adopted. As the accounts are associated with a specific type of malicious activity, besides computing the individual similarity, we compute and analyse pairwise cosine similarity among the accounts associated with malicious activities. Assume a dataset $D_a$ of malicious and benign accounts in a permissionless blockchain. We assume that each account is attributed to one and only one malicious activity. In reality, an account can have multiple associations. Consider, two malicious activities, $M_1$ and $M_2$ from a set of malicious activities $M$ that have set of accounts $A_{M_1}$ and $A_{M_2}$ associated with them, respectively. We compute cosine similarity ($CS_{i,j}$) such that $i\in A_{M_1}$, $j\in A_{M_2}$ and $A_{M_1}\bigcap A_{M_2}=\emptyset$ and then identify the probability of it being more than or equal to 0 ($p(CS_{i,j}\geq 0)$). If for all $i$ and $j$, $CS_{i,j}\geq 0$ then we say that the two malicious activities, $M_1$ and $M_2$, are similar.

Then, we use clustering algorithm with the motivation that accounts which are associated with the  same malicious activity would cluster together and show homophily~\cite{mdC2011}. We use K-Means algorithm to validate if indeed similarity exists between thee two malicious activities. Here, we assume an upper limit on $k$ (hyperparameter for K-Means) and use $k=||M||+1$. Note that $||M||$ represents the size of the set of different malicious activities, i.e., the number of different activities under consideration and $+1$ part represents the benign cluster. However, our results show that most of the accounts, even if they are associated with different malicious activities, cluster together. Note that in terms of the number of clusters found, the best case scenario would have been malicious accounts associated with different malicious activities cluster separately and the worst case would be all malicious accounts, irrespective of their associated malicious activity, cluster together.

\subsection{Bias Analysis}

The distribution of the number of accounts associated with each $M_i\in M$ is not uniform. This increases the sensitivity of the ML algorithm towards the $M_i\in M$ that is more prominent, i.e., has more number of associated accounts. Thereby, they induce a bias in the selected model towards $M_i$ that has the most number of associated accounts. To understand the effect of the number of malicious accounts attributed to a particular $M_i$ on the ML algorithm, we segment $D_a$ into different training and testing sub-datasets and use them to train and test ML algorithms. Let the set of different training and testing sub-datasets be $C=\{C_0,$ $C_1,\cdots,$ $C_n\}$ where each element $C_i$ represent a specific split of $D_a$. Let $Tr^{C_i}$ denote the training dataset, which contains 80\% of randomly selected accounts from $C_i$ and $Ts^{C_i}$ denote the test dataset, which contains the remaining 20\% accounts. The different $C_i$'s we use are:

\begin{itemize}
\item \textbf{Null model or $C_0$}: This is our baseline sub-dataset. Here we do not distinguish between the types of malicious activities rather only consider if an account is malicious or not. Note that here based on above notations, the training dataset is represented as $Tr^{C_0}$ and  the testing dataset as $Ts^{C_0}$.
\end{itemize}

Let ${A}_{M_1}^{S_0}$ represent the set of accounts associated with $M_1$ type malicious activity in the testing dataset, $Ts^{C_0}$. As our aim here is to analyze the bias caused due to a malicious activity, for example $M_1$, we analyse the results obtained when training and testing the ML algorithm using different combinations of accounts associated with $M_1$ activity. For instance, we analyse the performance of an ML algorithm when accounts associated with $M_1$ are not present in training dataset but are present the testing dataset. Below we list all such combinations that we use: 

\begin{itemize}
\item \textbf{$C_1$}: Here we train on $Tr^{C_0}$ but test on $Ts^{C_1}$ where $Ts^{C_1}=Ts^{C_0}-{A}_{M_1}^{S_0}$, i.e., we train on the 80\% of the dataset, but we remove all the accounts associated with activity $M_1$ from the testing dataset. Ideally, in this case, ML algorithm should perform similar to $C_0$ since the training dataset is same.

\item \textbf{$C_2$}: Here, we train on $Tr^{C_2}=Tr^{C_0}-{A}_{M_1}^{S_0}$ but test on $Ts^{C_2}$ which is same as $Ts^{C_0}$, i.e., we remove all the accounts associated with activity $M_1$ from the training dataset, but we keep the accounts associated with $M_1$ in the testing dataset. Ideally, in this case, ML algorithm should misclassify accounts associated to $M_1$ that are present in $Ts^{C_2}$. In case adverse results are obtained, it would mean that there is a bias.  

\item \textbf{$C_3$}: Here, we train on $Tr^{C_2}$ and test on $Ts^{C_3}$ which is same as $Ts^{C_1}$, i.e., we remove all the accounts associated with activity $M_1$ from both the training and the testing dataset. Ideally, in this case, ML algorithm should perform similar to $C_0$ since no accounts associated with $M_1$ activity are present in $Tr^{C_2}$ and $Ts^{C_3}$. In case adverse results are obtained, it would mean that there is a bias.  

\item \textbf{$C_4$}: Here, we train on $Tr^{C_4}=Tr^{C_0}+{A}_{M_1}^{S_0}$ and test on $Ts^{C_4}$ which is same as $Ts^{C_1}$, i.e., we remove all the accounts associated with activity $M_1$ from the testing dataset and add them to the training dataset. Ideally, in this case, ML algorithm should perform similar to $C_1$ since no accounts associated with $M_1$ activity are present in $Ts^{C_4}$.

\end{itemize}

Note that the above four configurations do test different, yet all, scenarios that are required to understand the effect of a malicious activity, $M_1$, on the ML algorithm. For the sake of completeness, we also consider the following training and testing sub-datasets:
\begin{itemize}

\item \textbf{$C_5$}: Here, we perform a 80-20 split of the number of accounts associated to each malicious activity in $M$. We then collect all these 80\% data-splits along with 80\% benign accounts to create the training dataset, $Tr^{C_5}$. Similarly, we collect all the remaining 20\% splits to create the testing dataset, $Ts^{C_5}$. We then train on the resulting $80\%$ of the dataset and test on the remaining $20\%$ of the dataset. Ideally, in this case, ML algorithm should perform similar to $C_0$. 

\end{itemize}

Among the supervised ML algorithms, in~\cite{agarwal2020detecting}, the authors presented that the ExtraTrees Classifier (\textbf{ETC}) performs the best on the data they had. We use ETC with the identified hyperparameters on the above-mentioned sub-datasets to identify the bias induced by a malicious activity. Further, as mentioned before, it is possible that a classifier trained on ETC might fail to capture new or adversarial data. Thus, we also apply different supervised ML algorithms on $C_0$ and identify the algorithm that achieves the best recall on the entire malicious class. We then execute the best identified ML algorithm on the above mentioned sub-datasets. To test the performance of different supervised ML algorithms on any new data, we collect the new and more recent malicious accounts transaction data ($D_b$) and execute the identified algorithms on the new transaction data. 

\subsection{Adversarial Analysis}

The new collected data ($D_b$) shows a low similarity with existing malicious activities. Such data does not classify as adversarial. We use CTGan~\cite{xu2019} to generate adversarial data for the malicious activities and use this new adversarial data ($D_g$) to validate our findings. Here, we use $D_g$ only on the test dataset, i.e., we perform training on $Tr^{C_0}$ while we perform our tests on $Ts^{D_g}$ that includes $D_g$ and all the benign accounts in $Ts^{C_0}$. Further, we also perform tests after training different ML algorithms when 1\%, 5\%, and 80\% of such adversarial feature vectors are added to the training dataset.

\begin{figure*}
    \center
    \includegraphics[width=0.75\textwidth]{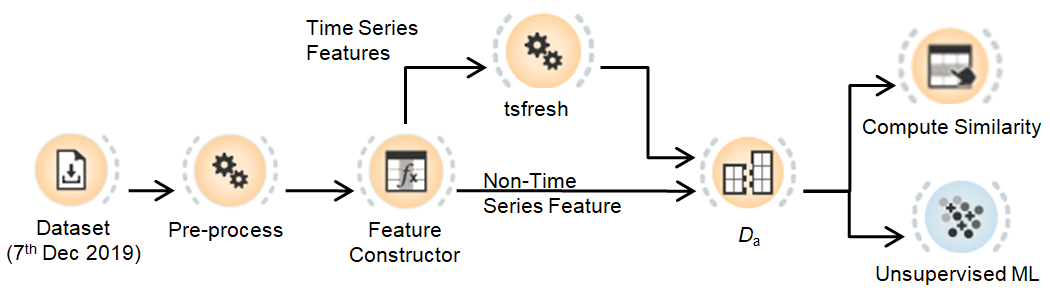}
    \caption{Overview of our Methodology used towards R1.}
    \label{fig:pipelineR1}
\end{figure*}
\begin{figure*}
    \center
    \includegraphics[width=0.8\textwidth]{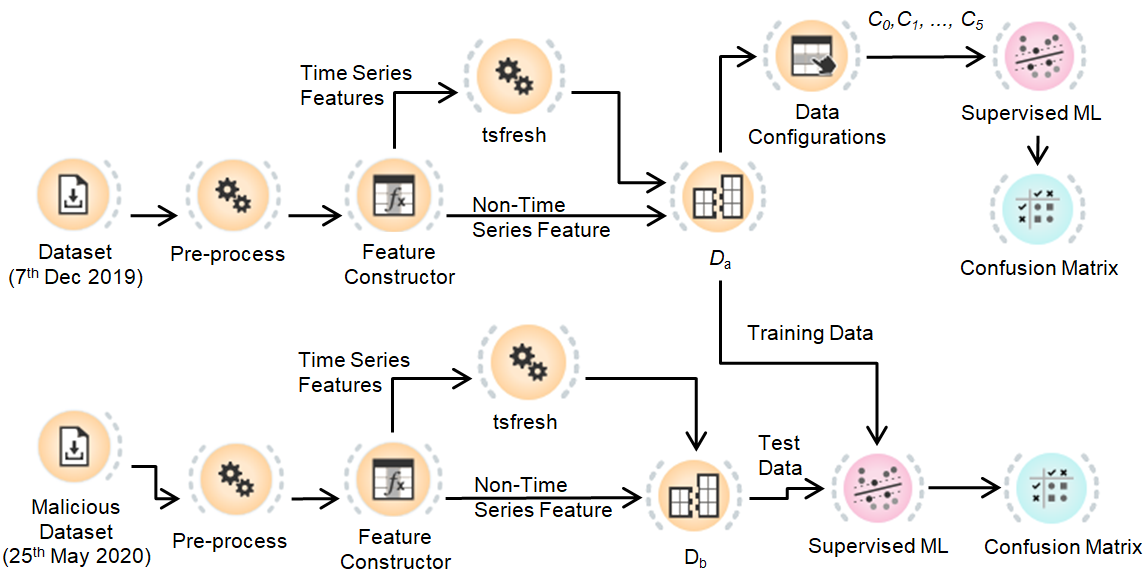}
    \caption{Overview of our Methodology used towards R2.}
    \label{fig:pipelineR2}
\end{figure*}
\begin{figure*}
    \center
    \includegraphics[width=0.7\textwidth]{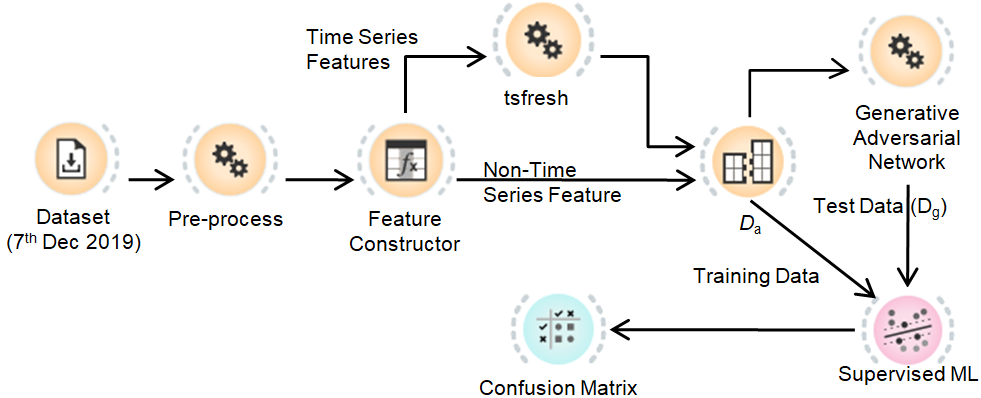}
    \caption{Overview of our Methodology used towards R3.}
    \label{fig:pipelineR3}
\end{figure*} 

In summary, Table~\ref{tab:notations} provides a list of notations we use. Here, note that we first generate the feature vector dataset using the same approach as listed by the authors in~\cite{agarwal2020detecting}. Then, towards methodology one \textbf{\textit{(R1)}}, we compute the cosine similarity amongst the malicious accounts in $D_a$ to understand how similar they are. We also apply unsupervised ML algorithms such as K-Means on the malicious accounts in $D_a$ to identify if malicious accounts cluster together (see Figure~\ref{fig:pipelineR1}). Towards methodology two \textbf{\textit{(R2)}}, we divide $D_a$ into different training and testing sub-datasets, $C=\{C_0,$ $C_1,\cdots,$ $C_5\}$, and execute different supervised ML algorithms to understand bias induced by a particular malicious activity and identify the performance of the reported classifier on the transaction data made available until 27th May 2020 (see Figure~\ref{fig:pipelineR2}). Towards methodology three \textbf{\textit{(R3)}}, we first filter out all the malicious accounts from $D_a$ and use CTGan separately on each malicious activity to generate adversarial ($D_g$) data for all the activities where we have more than 10 accounts. Note that there are two ways using which we can generate adversarial data: \textit{(1)} using feature vectors of malicious accounts and \textit{(2)} using features vectors of benign accounts. These represent, \textit{(1)} evading ML algorithm from being detected as malicious, and \textit{(2)} evading ML algorithms from being detected as benign and getting classified as malicious. In this work, we generate adversarial data using malicious accounts. Further, for CTGan, we use default parameters to synthesize the data, and use 1000 epochs (an epoch represents one iteration over the dataset) for fitting. We then compare the performance of various state-of-the-art supervised ML algorithms on $D_g$ (see Figure~\ref{fig:pipelineR3}).

\begin{table}[t]
    \centering
        \begin{tabular}{|r|l|}
        \hline
        \textbf{Notation}& {\textbf{Meaning}}\\
        \hline
        $D_a$& Dataset\\
        $M$ & Set of malicious activities present in $D_a$ \\
        $M_i$ & An element in $M$\\
        $A_{M_i}$ & Set of accounts in $D_a$ associated with $M_i$\\
        \multirow{2}{*}{$CS_{i,j}$} & Cosine similarity between accounts $i$\\
        & and $j$\\
        $p(CS_{i,j}\geq 0)$ & probability of $CS_{i,j}$ to be $>0$\\
        $k$ & Hyperparameter of K-Means algorithm\\
        $C$ & Set of sub-datasets of $D_a$\\
        $C_i$ & an element in $C$\\
        $Tr^{C_i}$ & Training dataset created from $C_i$\\
        $Ts^{C_i}$ & Testing dataset created from $C_i$\\
        \multirow{2}{*}{$A_{M_1}^{S_1}$} & Set of accounts associated with $M_1$ and\\
        & in $Ts^{C_1}$\\
        $D_b$ & New malicious accounts dataset\\
        $D_g$ & Adversarial dataset\\
        \multirow{2}{*}{$Ts^{D_g}$} & Testing dataset created from $D_g$ and \\
            & benign accounts in $Ts^{C_1}$.\\
        $F$ & Feature Vector\\
        \hline
          
        \end{tabular}
    \caption{List of Notations used in our work.}
    \label{tab:notations}
\end{table}

\section{Evaluation}~\label{sec:eval}

In this section, we first present the data we use and then provide a detailed analysis of our results.

\subsection{Data Used}

We use external transactions data present in the Ethereum main-net blockchain.
Ethereum~\cite{Vitalik2018} is one of the most widely adopted permissionless blockchain platform. It uses Proof-of-Work (PoW) consensus mechanism to validate transactions of the users. Ethereum provides users with the functionality to deploy additional programs called smart contracts which can be used to control the flow of the Ethers. In Ethereum, EOAs are accounts/wallet which is owned by a real-entity or a person; wherein the hash of public key of the owner of an EOA is the address of the EOA. On the other hand, SCs are similar to EOAs, with the exception that they contain code to automate certain tasks, such as sending and receiving Ethers, invoking, creating, and destroying other smart contracts when needed. SCs can be created and invoked both by EOAs and by other SCs. There are two types of transactions which occur on the Ethereum blockchain, \textit{External} and \textit{Internal} Transactions. While the external transactions occur between different EOAs and between EOAs and SCs, they are recorded on the blockchain ledger. Internal transactions are created by and occur between different SCs and are not recorded on the blockchain ledger. Further, SCs can execute external calls which are then recorded on the blockchain ledger. An external transaction typically has information about blockHash (hash of the block in which the transaction is present), blockNumber (another identifier to represent the block in which the transaction is present), the account from which the transaction is invoked, the account to which the transfer was made, gas (the amount `account present in the \textit{from} field of transaction' is willing to pay to the miners to include the transaction in the block), gasPrice (cost per unit gas), Transaction hash (hash of the transaction), balance (Ethers transferred), and the timestamp of the block. Such transaction are then grouped together into blocks before being published onto the blockchain.

We use Etherscan API~\cite{etherscanApi} to get transaction data of $2946$ malicious accounts that were marked until 7th December 2019. As the number of benign accounts were more than $117$ Million account, we perform under-sampling to get external transactions of $680,314$ benign accounts. Note that these 680,314 benign accounts and 2946 malicious accounts collectively constitute our dataset $D_a$ (defined in section~\ref{sec:method}). Since using a similar approach in~\cite{agarwal2020detecting}, the authors obtained good results on similar dataset by including the temporal features, we use the same set of $59$ features ($F$) in our work, as well. In short, these $59$ features are based on and can be classified under: \textit{(a)} temporal graph based features that includes indegree, outdegree, attractiveness, inter-event time burst, degree burst, and clustering coefficient, and \textit{(b)} transaction based features that includes activeness, crypto-currency balance and fee paid.

Etherscan provides for each malicious account a caution warning message so that other accounts can exercise caution while transacting with them. Other than such warning message, Etherscan also provides information about what malicious activity the account is involved in. These activities are: Phishing ($2590$ accounts), Scamming ($168$ accounts), Compromised ($21$ accounts), Upbit Hack ($123$ accounts), Heist ($13$ accounts), Gambling ($8$ accounts), Spam Token ($10$ accounts), Suspicious ($4$ accounts), Cryptopia Hack ($3$ accounts), EtherDelta Hack ($1$ accounts), Scam ($1$ accounts), Fake ICO ($2$ accounts), Unsafe ($1$ accounts), and Bugs ($2$ accounts). Thus, in the dataset, there are in total $14$ different types of malicious activities. For our different training and testing sub-datasets, we currently only focus on ``Phishing'' as it is the most prominent malicious activity in our dataset. Further, note that all the `Bitpoint Hack' accounts were also tagged under `Heist'. Therefore, we only use `Heist'. In $D_a$, we observe that $101$ unique EOAs created $153$ different malicious SCs. These EOAs are not marked as malicious in the dataset. Most of the EOAs created only one malicious SC, while, one EOA created $15$ malicious SCs. There are only $3$ SCs that are created by $3$ different SCs which in turn were created by the $2$ different EOAs. However, we refrain from adding these accounts in our analysis so as to not change any ground truth. We do not reveal the identities of these accounts because we do not want to malign any future transactions.  

As this list is dynamic, between 8th December 2019 and 27th May 2020, there were 1249 more accounts that were identified as malicious by Etherscan. On top, 487 accounts out of 2946 previously identified malicious accounts continued to transact until 27th May 2020. These total 1736 malicious accounts constitute our dataset $D_b$ (defined in section~\ref{sec:method}). The accounts in $D_b$ are associated with: Upbit Hack ($691$ accounts), Parity ($131$ accounts), Phishing ($842$ accounts), Ponzi ($38$ accounts), Gambling ($28$ accounts), Compromised ($1$ accounts), Unknown ($2$ accounts), Lendf.Me Hack ($2$ accounts), Plus Token Scam ($1$ accounts), Heist ($1$ accounts). We again notice that, in this case also, some accounts have more than 1 label associated with them. Based on our assumption, to ease our process, we associate these accounts with only one type of malicious activity. When analyzing $D_b$, we remove from $Tr^{C_0}$ all those accounts that were common in $D_a$ and $D_b$ and moved them to $Ts^{C_0}$, and retrained the different supervised ML algorithms to identify their performance.

\subsection{Results}

All our results are averaged over 50 iterations on our dataset and are generated using Python3.

\begin{figure}
    \includegraphics[width=\columnwidth]{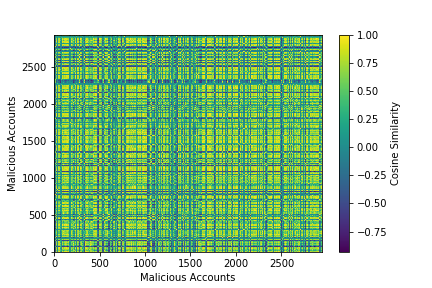}
    \caption{Cosine similarity between malicious accounts.\label{fig:csMalVSMal}}
\end{figure}

\subsubsection{R1: Similarity Analysis}

\begin{figure*}
    \centering
    \subfloat[$p(CS_{i,j}<0)$][$p(CS_{i,j}<0)$]{
        \includegraphics[width=\columnwidth]{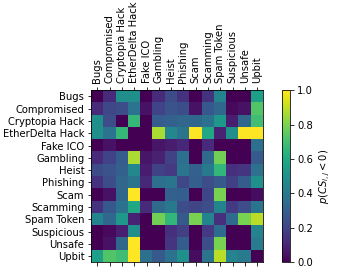}
        \label{fig:csfirstBin}
    }
    \subfloat[$p(CS_{i,j}\geq0)$][$p(CS_{i,j}\geq0)$]{
        \includegraphics[width=\columnwidth]{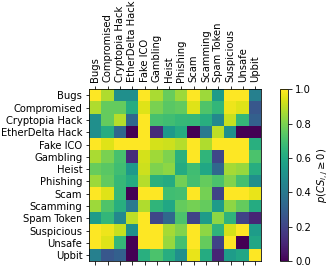}
        \label{fig:cssecondBin}
    }
        \caption{Probability that accounts associated with two malicious activities show similarity more than 0 or less than 0.}\label{fig:csTagBased}
\end{figure*}

Some of the malicious activities present in the dataset have similar definition. We calculate cosine similarity score between all the malicious accounts present in $D_{a}$ to identify similarity that exists between them (see Figure~\ref{fig:csMalVSMal}). From the figure, we observe that in many cases $CS_{i,j}<0$, as expected because they potentially belong to different malicious activities. Nonetheless, we also observe that some pair of accounts have high similarity between them. There are two reasons for such high similarity: \textit{(1)} these accounts could actually belong to the same malicious activity, and \textit{(2)} although, the account might represent same malicious activity, in Ethereum, they might be marked differently, i.e., these accounts have been marked for malicious activities that have similar description. To understand this further, we check all the accounts associated with a pair of malicious activities and mark the two activities as similar if all the accounts in one type of malicious activity have cosine similarity more than 0 with all the malicious accounts in the other type. Figure~\ref{fig:csTagBased} depicts the probabilities $p(CS_{i,j}<0)$ (see Figure~\ref{fig:csfirstBin}) and $p(CS_{i,j}\geq 0)$ (see Figure~\ref{fig:cssecondBin}) between all the accounts related to two malicious activities. Note that the two figures are complementary to each other as $p(CS_{i,j}<0)=1-p(CS_{i,j}\geq 0)$. From the figures, we notice that many activities show high probability of similarity with other malicious activities, for example, `Bugs' and `Unsafe'. There are~$\approx$ 158 pairs of malicious activities where $p(CS_{i,j}\geq 0)>0.5$. Further, we also notice that within `Phishing', although most account are similar, there are some accounts which show dissimilarity. 

We use K-Means algorithm to analyze the clustering patterns of all malicious accounts present in the dataset $D_a$ to further understand if the accounts identified as similar show homophily. Here, we use $k=15$. We see that most of the malicious accounts, irrespective of the malicious activity they belong to, cluster together in at most 3 clusters except for the accounts that belong to malicious activities `Heist', `Compromised', and `Gambling' (see Table~\ref{tab:kmeans}). Further, in the cluster that had most number of the malicious accounts (i.e., cluster \#1), except for accounts associated with `Upbit Hack' and `Spam Token', all other malicious accounts clustered together in cluster \#1.

Therefore, we infer that different malicious activities in a blockchain such as Ethereum, behave in a similar manner. Same labels could be used to depict certain malicious activities, such as `Phishing' and `Fake ICO'. Currently, we do not need 14 different labels as most of the malicious activities could be captured in at most 3 clusters. 

\begin{table}[t]
    \centering
    \begin{threeparttable}
        \begin{tabular}{|c|c|c|c|c|}
            \textbf{Tag Name}& \textbf{Total}& \rot{\textbf{Cluster 1 }} & \rot{\textbf{Cluster 2 }} & \rot{\textbf{Cluster 3 }}\\
            \hline
            Phishing & 2590 & 1700  & 492  & 398 \\
            Upbit Hack$\dagger$ & 123 & 32  & 1 & 90 \\
            Scamming & 168 & 116  & 27 & 25  \\
            Heist$\ddagger$ & 13 & 9  & 1 & 2  \\
            Compromised$\ddagger$ & 21 & 17  & - & 1 \\
            Unsafe & 1 & 1 & - & - \\
            Spam Token$\dagger$ & 10 & 1  & 3 & 6 \\
            Bugs & 2 & 2 & - & -  \\
            EtherDelta Hack & 1 & - & 1 & -\\
            Cryptopia Hack & 3 & 2 & - & 1 \\
            Gambling$\ddagger$ & 8 & 7  & -  & - \\
            Suspicious & 4 & 3  & - & 1  \\ 
            Fake ICO & 2 & 2 & - & -  \\
            Scam & 1 & 1 & - & - \\
            \hline
        \end{tabular}\begin{tablenotes}
            \item [$\dagger$] not well clustered in cluster 1, $\ddagger$ not well clustered in 3 clusters 
        \end{tablenotes}
    \end{threeparttable}
    \caption{Clustering of malicious account using K-Means. Here clusters are ranked based on the number of malicious accounts they cluster.}
    \label{tab:kmeans}
\end{table}

\subsubsection{R2: Bias Analysis}

\begin{table}[t]
    \centering
    \begin{threeparttable}
        \begin{tabular}{|c|c|c|c|c|c|}%
            & & \multicolumn{2}{c|}{\textbf{ETC}} & \multicolumn{2}{c|}{\textbf{NN}}\\
            \cline{3-6}
            \textbf{Data}&&\multicolumn{2}{c|}{\textbf{Recall}} & \multicolumn{2}{c|}{\textbf{Recall}} \\
            \cline{3-6}
            \textbf{Until}&\textbf{sub-datasets}&\textbf{Mal} & \textbf{Ben}& \textbf{Mal} & \textbf{Ben} \\
            \hline
            \multirow{6}{*}{\rot{07/12/2019}}&$C_0$ 
            & 0.70 
            & 0.99 &0.83 &0.94 \\
            &$C_1$ 
            & 0.76 
            & 0.99 &0.86 &0.89 \\
            &$C_2$ 
            &0.28 
            &0.99 &0.79 &0.93\\
            &$C_3$ 
            & 0.59 
            &0.99 &0.83 &0.97 \\
            &$C_4$ 
            & 0.79 
            &0.99 &0.87 &0.90\\
            &$C_5$ 
            & 0.70 
            & 0.99 &0.88 &0.91\\
            \hline
        \end{tabular}
        \begin{tablenotes}
            \item [ETC]  \small{ExtraTreesClassifier(class\_weight = `balanced', criterion = `entropy', max\_features = 0.3, max\_samples = 0.3, min\_samples\_leaf = 14, min\_samples\_split = 20, n\_estimators = 200)}
            \item [NN] \small{NeuralNetworks(epochs = 50, regularization = l2(0.0001), dropout = 0.5,loss=`binary crossentropy', optimizer=`adam' )}
        \end{tablenotes}
    \end{threeparttable}
    \caption{Recall achieved using ExtraTrees Classifier and Neural Network for different sub-dataset. Here `Mal' represents malicious class and `Ben' represents the benign class.}
    \label{tab:cmSplits}
\end{table}

We first test ETC, as~\cite{agarwal2020detecting} reported it to produce the best results, on different training and testing sub-datasets ($C_i$) to understand the bias induced due to the imbalance in the number of accounts associated with a particular malicious activity. For ETC, we use the hyperparameters reported by~\cite{agarwal2020detecting}. These are: class\_weight = `balanced', criterion = `entropy', max\_features = 0.3, max\_samples = 0.3, min\_samples\_leaf = 14, min\_samples\_split = 20, n\_estimators = 200. All other hyperparameters are kept as default. In our dataset, since most number of malicious accounts are associated with `Phishing' activity, here we choose `Phishing' for our study and our training and testing sub-datasets are configured accordingly. Note that this analysis is also valid for other malicious activities. From Table~\ref{tab:cmSplits}, we observe that, for $C_2$ and $C_3$, the recall on the accounts tagged as malicious deteriorates significantly. For $C_2$, we expected such results because there are no `Phishing' accounts in the training dataset ($Tr^{C_2}$), but are present in test dataset ($Ts^{C_0}$).  However, for $C_3$, the recall on the malicious accounts was below expectation. This proves the existence of bias in ETC towards `Phishing'. To understand further and know which all malicious activities are impacted due to such bias, we study the confusion matrix and the distribution of malicious accounts therein. 

\begin{table*}[t]
    \centering
    \begin{threeparttable}
        \begin{tabular}{|c|c|c|c|c|c|c|c|c|c|c|c|c|c|c|}
        \hline
         & \multicolumn{2}{c|}{\textbf{Total}} & \multicolumn{6}{c|}{\textbf{ETC}} & \multicolumn{6}{c|}{\textbf{NN}} \\
        \hline
\multirow{2}{*}{\textbf{Activity}} &\textbf{$Ts^{C_0}$..}& \multirow{2}{*}{$Ts^{C_5}$} & \multirow{2}{*}{\textbf{$Ts^{C_0}$}}          & \multirow{2}{*}{\textbf{$Ts^{C_1}$}}          & \multirow{2}{*}{\textbf{$Ts^{C_2}$}}           & \multirow{2}{*}{\textbf{$Ts^{C_3}$}}         & \multirow{2}{*}{\textbf{$Ts^{C_4}$}}          & \multirow{2}{*}{\textbf{$Ts^{C_5}$}}          & \multirow{2}{*}{\textbf{$Ts^{C_0}$}}          & \multirow{2}{*}{\textbf{$Ts^{C_1}$}}          & \multirow{2}{*}{\textbf{$Ts^{C_2}$}}           & \multirow{2}{*}{\textbf{$Ts^{C_3}$}}         & \multirow{2}{*}{\textbf{$Ts^{C_4}$}}          & \multirow{2}{*}{\textbf{$Ts^{C_5}$}} \\

&\textbf{$..Ts^{C_4}$}& &          &  &  &  &  &  &      &          & & & & \\
\hline
Phishing   &517 &   508 & 359        & -        & 123           & -         & -        & 352        & 432          & -          & 404           & -         & -        & 442     \\ 
Upbit  &18  & 25  & 17         & 17          & 17          & 17         & 17       & 25        & 17        & 18          & 17          & 17         & 18         & 25     \\ 
Scamming &33 &  34  & 22         & 23         & 15$\dagger$          & 15$\dagger$         & 23         & 23         & 26         & 27         & 24$\ddagger$          & 24$\ddagger$         & 27         & 28      \\ 
Heist & 1 & 3 & 0          & 0          & 0           & 0         & 0          & 1          & 1          & 1          & 1           & 1         & 1          & 2      \\ 
Compro.. &5 & 5  & 5          & 5          & 4           & 4         & 5          & 4          & 2          & 2          & 5           & 4         & 4          & 5     \\ 
Unsafe &1 &1  & 0          & 0          & 0           & 0         & 0          & 0          & 1         & 1          & 1           & 1         & 1          & 1      \\ 
Spam T.. & 2 & 2 & 0          & 0          & 0           & 0         & 2          & 1          & 0          & 2          & 2           & 2        & 2          & 2      \\ 
Bugs &-  & 1 & -          & -          & -           & -         & -          & 1          & -          & -          & -           & -         & -          & 1       \\ 
EtherDelta &-  & 1 & -          & -          & -           & -         & -          & 0          & -          & -          & -           & -         & -          & 1         \\ 
Cryptopia &- & 1 & -          & -          & -           & -         & -         & 1          & -          & -          & -           & -         & -          & 1       \\ 
Gambling &3  &2 & 2          & 2          & 2           & 1         & 2          & 0          & 2          & 2          & 2           & 2         & 1          & 2         \\ 
Suspicious &- & 1 & -          & -          & -           & -         & -          & 0         & -          & -          & -           & -         & -          & 1        \\ 
Fake ICO &1 & 1 & 1          & 1          & 0 $\bullet$           & 0 $\bullet$         & 1          & 1          & 1          & 1           & 1   $\odot$         & 1   $\odot$     & 1          & 1        \\
\hline
\end{tabular}
    \end{threeparttable}
    \begin{tablenotes}
    \item $\dagger$ number of `Scamming' accounts  correctly classified by ETC is less $\ddagger$  number of `Scamming' accounts  correctly classified by NN is more
    \item  $\bullet$ no `Fake ICO' account correctly classified $\odot$ NN classifies accounts associated with `Fake ICO' correctly
    \item `-' symbol represents that accounts associated with the particular malicious activity were not present in our test set
    \item `Spam T..' represents accounts associated with `Spam Token' ; `Compro..' represents accounts associated with `Compromised'
    \end{tablenotes}
    \caption{Malicious activity based analysis of different confusion matrices for ETC and NN. Note that here `Scam' activity is not present as the only associated account was selected in the Training dataset.}
    \label{tab:accountInCMSplit}
\end{table*}

From the Table~\ref{tab:accountInCMSplit}, we observe that for $C_2$, though only accounts attributed to `Phishing' activities are removed from the training dataset, more than 50\% of the accounts attributed to `Scamming' are misclassified. Same results are observed for $C_3$, when accounts associated with `Phishing' are removed from both the training dataset ($Tr^{C_3}$) and the test dataset ($Ts^{C_3}$). However, if `Phishing' related accounts are present in the training dataset, as in $C_0$, $C_1$ and $C_4$, less number of accounts tagged as `Scamming' are misclassified. A similar observation is made for the accounts associated with `Fake ICO'. This is also obtained from the results in the previous subsection (R1), where we see that `Phishing' and `Scamming' based accounts, show high similarity. This validates the results we obtain using different configurations.

\begin{table}[t]
    \centering
    \begin{threeparttable}
        \begin{tabular}{|c|c|c|c|c|c|c|}
        \hline
            \textbf{ML}& \multicolumn{2}{c|}{\textbf{$Ts^{C_0}$}}& \multicolumn{2}{c|}{\textbf{$D_b$}}& \multicolumn{2}{c|}{\textbf{$D_g$}} \\
            \cline{2-7}
             \textbf{algorithm}& \textbf{Mal} & \textbf{Ben} & \textbf{Mal} & \textbf{Ben} & \textbf{Mal} & \textbf{Ben}\\
            \hline
            RF & 0.66 & 0.99 & 0.0 & 0.99 & 0.54 & 0.99\\
            XG Boost & 0.70 & 0.99 & 0.003 & 0.99 & 0.60 & 0.99 \\
            DT & 0.71 & 0.99 & 0.04 & 0.98 & 0.67 & 0.99\\
            ADaBoost & 0.51 & 0.99 & 0.001 & 0.99 & 0.56 & 0.99\\
            Grad. Boost. & 0.72 & 0.99 & 0.006 & 0.99 & 0.38 & 0.99 \\
            ETC & 0.70 & 0.99 & 0.0 & 0.99 & 0.70 & 0.99 \\
            \textbf{NN} & \textbf{0.83} & \textbf{0.94} & \textbf{0.25} & \textbf{0.89} & \textbf{0.95} & \textbf{0.90} \\
            \hline
        \end{tabular}
    \end{threeparttable}
    \caption{Recall score obtained when different ML algorithms are trained on $Tr^{C_0}$. Here `Mal' represents malicious class, `Ben' represents the benign class, `RF' represents RandomForest classifier, `DT' represents Decision Trees, `Grad. Boost.' represent Gradient Boosting, `ETC' represents ExtraTrees classifier, and `NN' represent Neural Networks.}
    \label{tab:SupMlComparison}
\end{table}

Since ETC shows bias, we execute different supervised ML algorithms on the different training and test sub-datasets created from $D_a$. Here, on the features we have, we study those ML algorithms that were used by the related works. Table~\ref{tab:SupMlComparison} depicts the recall obtained after different supervised ML algorithms were studied on $C_0$. Here, we keep the hyperparameters of the supervised ML algorithms to either default or to those values that are reported by the related works. Apart from the supervised ML algorithms specified in Table~\ref{tab:SupMlComparison}, we also test the performance of GCN using our features on the dataset $D_a$. GCN achieves a recall score of 32.1\% on the dataset $D_a$. From the set of  ML algorithms, we identify that Neural Network (NN) (with hyperparameters: epochs = 50, regularization = l2(0.0001), dropout = 0.5, loss=`binary crossentropy', optimizer=`adam') performs the best on $C_0$ and achieves a recall $0.83$ on the malicious class. However recall on benign class drops to $0.94$ resulting in a balanced-accuracy (average recall on all classes) of $88.5\%$. We also note that although recall on malicious class is high for NN, recall on benign class is relatively lower. Overall, the balanced accuracy is still high. To understand if NN is biased towards `Phishing', we execute NN on different training and test sub-datasets we have. We find that NN has better recall on $C_2$ and $C_3$ as well (see Table~\ref{tab:cmSplits}). Further, we find that NN was able to correctly classify most of the accounts associated with different malicious activities irrespective of the bias (see Table~\ref{tab:accountInCMSplit}).

\begin{figure}
     \includegraphics[width=\columnwidth]{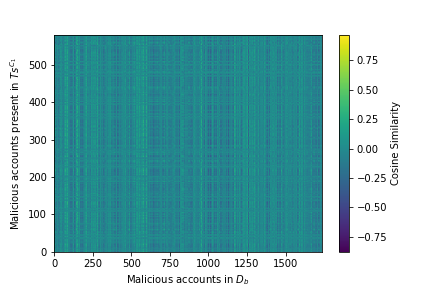}
     \caption{Cosine similarity between malicious accounts present in $D_b$ and $Ts^{C_0}$. \label{fig:maltestDB}}
 \end{figure}

Next, we perform experiments to understand which supervised ML algorithm performs the best in the scenario when new data on already known malicious activities is presented. From Figure~\ref{fig:maltestDB}, we find that the similarity between the malicious accounts in $Ts^{C_0}$ and those in $D_b$ is less. We then test the performance of different supervised ML algorithms using $D_b$ to understand if these new malicious accounts are classified correctly on already trained supervised ML algorithms, i.e., when we train on $Tr^{C_0}$ while test on all benign accounts in $TS^{C_0}$ and all malicious account in $D_b$. Table~\ref{tab:SupMlComparison} also presents the recall score obtained in this scenario. Here, we again observe that Neural Network performs better than the rest of the supervised ML algorithms and was able to classify more than $434$ account out of $1736$ malicious accounts. Thus, we get further confirmation that NN performs the best and is not affected by the bias caused by high number of accounts associated with `Phishing' malicious activity in the Ethereum transaction data.

\subsubsection{R3: Adversarial Analysis}

Next, to answer our other research question and test the effect of adversarial data on the supervised ML algorithms, we first generate adversarial data from accounts attributed to the particular malicious activities and then test various supervised ML algorithms on the generated datasets ($D_g$).
For each malicious activity where the number of associated accounts are sufficiently large, we generate $1000$ adversarial samples. For malicious activities that are moderately represented, we generate 50 adversarial samples. Since for all the other malicious activities we had less than 10 accounts, we did not generated adversarial accounts for those malicious activities. Thus, in $D_g$, the distribution of accounts associated with different malicious activities is as follows: Phishing ($1000$ accounts), Scamming ($1000$ accounts), Upbit Hack ($1000$ accounts), Spam Token ($50$ accounts), Compromised ($50$ accounts), Heist ($50$ accounts). Here, the number of accounts generated for each malicious activity are in accordance to the number of malicious accounts available for each activity. As for `Phishing', `Scamming', and `Upbit Hack' the number of accounts were more, we generated 1000 each adversarial feature vectors. This number is randomly selected and is based on the computational power available to us. 

\begin{figure*}
    \centering
    \subfloat[Cosine similarity between malicious accounts in $D_a$ and those in $D_g$][Cosine similarity between malicious accounts in $D_a$ and those in $D_g$]{
        \includegraphics[width=\columnwidth]{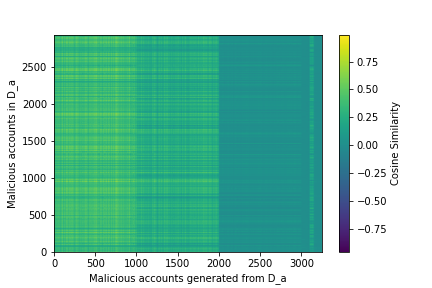}
        \label{fig:MAlvsGAN}
    }
    \subfloat[Cosine similarity between phishing accounts in $D_a$ and those in $D_g$][Cosine similarity between phishing accounts in $D_a$ and those in $D_g$]{
        \includegraphics[width=\columnwidth]{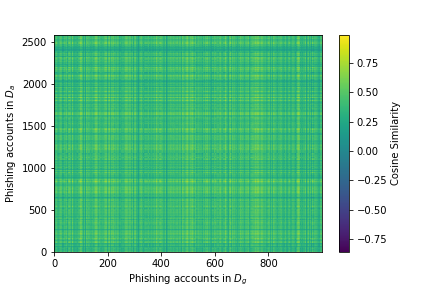}
        \label{fig:phishgen}
    }
        \caption{Cosine Similarity Scores.}\label{fig:csGan}
\end{figure*}

As expected, $D_g$ has a high similarity with the source dataset (see Figure~\ref{fig:MAlvsGAN}) and can be considered as the dataset meant to evade state-of-the-art ML algorithms. From Figure~\ref{fig:MAlvsGAN}, we note that some pairs of accounts have low similarity. This is true as Figure~\ref{fig:MAlvsGAN} shows similarity between all malicious accounts which belong to different malicious activities. Nonetheless, if we see the similarity just between accounts associated with `Phishing' activity in $D_a$ and in $D_g$, we observe a similar behavior (see Figure~\ref{fig:phishgen}). Some accounts in $D_g$ show similarity as low as $-0.84$. This is true because some malicious accounts associated with Phishing in $D_a$ had low similarities amongst themselves (see Figure~\ref{fig:csfirstBin}). Since accounts in $D_g$ are based the accounts in $D_a$, this dataset also shows a same behavior. 

In this scenario, we observe that NN algorithm also achieved the best recall ($>94.8\%$) on the malicious class among the set of supervised ML algorithms (see Table~\ref{tab:SupMlComparison}). Comparing with the instance when adversarial data was not presented, we note that previously identified ML algorithms did not perform well. Their recall is much less than $70.1\%$. For GCN, however, as $D_g$ is a generated feature vector dataset, we did not have associated graph. Therefore, we were not able to perform any tests on GCN. Here, we infer that the identified NN model is more robust and is able to capture any adversarial data/attack. Note that here we did not train NN on any adversarial feature vectors. 

We expect to achieve better results when some adversarial feature vectors are provided in the training. To test the effect, we test of three different training dataset configurations: \textit{(1)} include 1\% randomly selected accounts from $D_g$, \textit{(2)} 5\% randomly selected accounts from $D_g$, and \textit{(3)} 80\% feature vectors of all the malicious activities in $D_g$. For sake of completeness, we test on supervised ML algorithms used in the related work. Table~\ref{tab:dginTrain} provides recall of both malicious and benign accounts, when different algorithms are trained on above-mentioned three configurations. Here, we note that RandomForest, for the default hyperparameter values, performs best in all the three cases. Here NN, does perform well but recall on both malicious and benign classes are better for RandomForest classifier. Further, note that although recall is less for malicious class when we perform tests considering 80\% feature vectors of all the malicious activities in $D_g$, but the balance accuracy is better than the other two cases.


\section{Conclusion and Discussions}\label{sec:conclusion}
With a rise in their adoption, permissionless blockchains are often victim to various cyber-attacks, scams, and host to ransom payments. Most of the security threats and malicious activities in permissionless blockchains exploit the vulnerabilities exposed due to platform, network architecture, programming bugs, and sometimes are socially engineered. Currently, different ML algorithms are used in the state-of-the-art techniques to detect such exploits. However, these techniques have limitations as they do not distinguish between different types of the exploits. 

\begin{table}[t]
    \centering
    \begin{threeparttable}
        \begin{tabular}{|c|c|c|c|c|c|c|}
        \hline
            \textbf{ML}& \multicolumn{2}{c|}{\textbf{1\%}}& \multicolumn{2}{c|}{\textbf{5\%}}& \multicolumn{2}{c|}{\textbf{80\% of all}} \\
            \cline{2-7}
             \textbf{algorithm}& \textbf{Mal} & \textbf{Ben} & \textbf{Mal} & \textbf{Ben} & \textbf{Mal} & \textbf{Ben}\\
            \hline
            RF & \textbf{0.94} & \textbf{0.99} & \textbf{0.99} & \textbf{0.99} & \textbf{1.00} & \textbf{0.99}\\
            XG Boost & 0.81 & 0.99 & 0.96 & 0.99 & 1.00 & 0.99 \\
            DT & 0.90 & 0.99 & 0.97 & 0.99 & 1.00 & 0.99\\
            ADaBoost & 0.61 & 0.99 & 0.89 & 0.99 & 0.98 & 0.99\\
            Grad. Boost. & 0.88 & 0.99 & 0.91 & 0.99 & 0.99 & 0.99 \\
            ETC & 0.80 & 0.99 & 0.93 & 0.99 & 1.00 & 0.99 \\
            \textbf{NN} & 0.92 & 0.96 & 0.97 & 0.87 & 0.93 & 0.96 \\
            \hline
        \end{tabular}
    \end{threeparttable}
    \caption{Recall score obtained when different ML algorithms are trained on adversarial feature vectors. Here `Mal' represents malicious class, `Ben' represents the benign class, RF represents RandomForest classifier, DT represents Decision Trees, Grad. Boost. represent Gradient Boosting, ETC represents ExtraTrees classifier, and NN represent Neural Networks.}
    \label{tab:dginTrain}
\end{table}

In this paper, we first study similarity among accounts to understand if presented ML models are biased towards a specific malicious activity. We identify that most of the malicious activities can be clustered in mostly 3 clusters and existing ML models are biased towards the malicious activity that has maximum number of accounts associated with it. We then identify that NN, despite such bias, was able to achieve much better recall on the malicious class and also detect any adversarial attacks. Thus, in the future, if any attacker is able to intelligently craft transactions that show adversarial behavior, they can be detected. We identify that the NN is robust to such adversarial attacks and is able to classify malicious account with high recall. Further, even if 1\% adversarial inputs are added to the training dataset, performance of NN increases by 1.5\% to achieve a balanced accuracy of 94\%. However, in this case RandomForest classifier performs the best among the set of supervised ML algorithms and achieves a balanced accuracy of 96.5\%. Nonetheless, the performance of NN (or any other supervised algorithm) could vary if any new features are introduced. Further, we obtained the labels representing malicious activities from Etherscan that uses crowd-sourced mechanism where anyone can suggest a label name for an account\footnote{Name Tagging In Etherscan: \url{https://etherscan.io/contactus}}. Such aspects cause diversity in the label names where these label names could refer to the same malicious activity. This raises questions upon the correctness of the label names. Nonetheless, one could use natural language processing (NLP) approaches to identify similar label names and group them accordingly. 

One future direction that we would like to pursue following this study is to analyse the behavior of accounts associated with new malicious activities in the permissionless blockchains. For this, we would like to integrate `Active Learning' into our ML pipeline to do real-time analysis of the accounts. Since, in our study, we do not focus on the internal transactions, we would also like to use them to detect transactions based anomalies such as those caused due to BZX exploit\footnote{BZX Exploit: \url{https://etherscan.io/accounts/label/bzx-exploit}}. Since internal transactions are carried out by smart contracts, studying them would help better understand the transactional behavior of the smart contracts, and thus detect those attacks as well where transactions utilize multiple SCs. As reported before, we observed that, in most cases, the creator (either an EOA or an SC) of the SC which is involved in malicious activity as well as the SC created by such malicious SC is not tagged as malicious. Therefore, we would like to analyse such accounts as well in future.

\section*{Acknowledgement}

We thank the authors of~\cite{agarwal2020detecting} for sharing with us the account hashes of all the 2946 malicious accounts until 7th December 2019, 680,314 benign accounts, and 1736 malicious accounts until 27th May 2020.

\section*{Availability}
All our codes along with list of malicious accounts used are available for re-use and test upon request.

\bibliographystyle{plain}
\bibliography{biblio.bib}

\appendix
\section{Appendix: Burst and Attractiveness}

This section provides a brief overview of the Burst and Attractiveness related features which we have used in our study:
 \begin{itemize}
     \item \textbf{\textit{Burst}}: An entity is said to exhibit \textit{bursty} behavior if it shows irregular changes in the sequence of events associated with it. Bursts for different entities is defined as:
     \begin{enumerate}
         \item \textbf{Degree Burst}: In graph theory, the degree of a vertex is defined as the number of edges it has. An entity is said to have a `Degree Burst' if the number of edges the entity has at a given epoch is greater than a threshold value. With respect to blockchains, such aspect captures irregularity shown by the malicious accounts, such as those involved in `Upbit Hack'.
         
         \item \textbf{Balance Burst}: 
         In our work, we say that a particular account has a `Balance Burst' if the crypto-currency involved in a transaction by the same account at a particular epoch is greater than a threshold value. Usually, accounts involved in malicious activities, such as money laundering, tend to transfer high amount of crypto-currency. Capturing such events would lead to identification of accounts showing malicious behavior, such as the accounts involved in Silk-Road activities~\cite{Spagnuolo2014}.
         
         \item \textbf{Transaction fee Burst}: An adversary can attempt to bribe a miner into accepting a particular transaction in a block by raising the fee corresponding to a transaction. Capturing burst of fees paid, would capture such instances. Therefore, `Transaction fee Burst' is the number of instances where the `Transaction fee' involved in a transaction is greater than a threshold value.
         
         \item \textbf{Inter-event time burst}: Inter-event time is defined as the time between two transactions. There can be instances where an account in a blockchain transacts with others such that the distribution of inter-event time is non-uniform. Such non-uniformity, leads to having multiple transactions in a short time period. Therefore, `Inter-event time burst' is defined to capture such aspects.  
         
     \end{enumerate}
     \item \textbf{\textit{Attractiveness}} : The accounts involved in malicious activities usually tend to carry out transactions with the new accounts overtime. Such accounts show less stability in terms of accounts they have transacted with before. `Attractiveness' for an account is thus the probability of an account receiving crypto-currency from a previously unknown accounts at a given epoch. 
 \end{itemize}
 
\section{Appendix: List of Features used}\label{app:b}
The set of 59 features used in our work is: 
$F=\{$indegreeTimeInv, 
outdegreeTimeInv, 
degreeTimeInv,
numberOfburstTemporalInOut, 
longestBurstTemporalInOut, 
numberOfburstTemporalIn, 
longestBurstTemporalIn, 
numberOfburstTemporalOut, 
longestBurstTemporalOut, 
numberOfburstDegreeInOut, 
longestBurstDegreeInOutAtTime, 
numberOfburstDegreeIn, 
longestBurstDegreeInAtTime, 
numberOfburstDegreeOut, 
longestBurstDegreeOutAtTime, 
zeroTransactions, 
totalBal, 
transactedFirst, 
transactedLast, 
activeDuration, 
averagePerInBal, 
uniqueIn, 
lastActiveSince, 
indegree\_\_index\_mass\_quantile\_\_q\_0.1,
indegree\_\_energy\_ratio\_by\_chunks\_\_num\_segments\_10\_\_-segment\_focus\_0, 
indegree\_\_linear\_trend\_\_attr\_``pvalue", 
ittime\_\_quantile\_\_q\_0.7, 
ittime\_\_fft\_coefficient\_\_coeff\_0\_\_attr\_``real", 
ittime\_\_median, 
outdegree\_\_energy\_ratio\_by\_chunks\_\_num\_segments\_10-\_\_segment\_focus\_0, 
outdegree\_\_enegy\_ratio\_by\_chunks\_\_-num\_segments\_10\_\_segment\_focus\_1, 
outdegree\_\_fft\_coefficient\_\_coeff\_0\_\_attr\_``real", 
gasPrice\_\_quantile\_\_q\_0.2, 
gasPrice\_\_quantile\_\_q\_0.1, 
gasPrice\_\_cwt\_coefficients\_\_widths\_(2, 5, 10, 20)\_\_coeff\_0\_\_w\_20, 
attractiveness\_\_median, 
attractiveness\_\_quantile\_\_q\_0\_0.4, 
attractiveness\_\_mean, 
balanceOut\_\_quantile\_\_q\_0.1, 
balanceOut\_\_quantile\_\_q\_0.3, 
balanceOut\_\_cwt\_coefficients\_\_widths\_(2, 5, 10, 20)\_\_coeff\_0\_\_w\_2, 
balanceIn\_\_quantile\_\_q\_0.4, 
balanceIn-$ $\_\_cwt\_coefficients\_\_widths\_(2, 5, 10,20)\_\_coeff\_0\_\_w\_20, 
balanceIn\_\_quantile\_\_q\_0.3, 
maxInPayment\_\_quantile\_\_q\_0.3, 
maxInPayment\_\_quantile\_\_q\_0.2, 
maxInPayment\_\_cwt\_coefficients\_\_widths\_(2, 5, 10, 20)\_\_coeff\_0\_\_w\_5, 
maxOutPayment\_\_quantile\_\_q\_0.6, 
maxOutPayment\_\_quantile\_\_q\_0.1, 
maxOutPayment\_\_cwt\_coefficients\_\_widths\_(2, 5, 10, 20)\_\_coeff\_0\_\_w\_2, 
clusteringCoeff, 
burstCount\_gasPrice, 
burstCount\_balanceIn, 
burstCount\_balanceOut, 
burstInstance\_indegree, 
burstInstance\_outdegree, 
burstInstance\_outdegree, 
burstInstance\_maxInPayment, 
burstInstance\_maxOutPayment, 
burstInstance\_gasPrice$\}$

\end{document}